\documentclass[%
 reprint,
 amsmath,amssymb,
 aps,
]{revtex4-2}

\usepackage{graphicx}
\usepackage{dcolumn}
\usepackage{bm}
\usepackage{epsfig}
\usepackage{float}
\usepackage{color}

\def\sideremark#1{\ifvmode\leavevmode\fi\vadjust{\vbox to0pt{\vss
 \hbox to 0pt{\hskip\hsize\hskip1em
 \vbox{\hsize2cm\tiny\raggedright\pretolerance10000
  \noindent #1\hfill}\hss}\vbox to8pt{\vfil}\vss}}}

\newcommand{\Z}{\mathbb Z}

\newcommand{\tens}{\otimes}
\newcommand{\extd}{{\rm d}}
\newcommand{\eps}{\varepsilon}
\renewcommand{\L}{\mathcal L}
\newcommand{\id}{{\rm id}}

\begin{document}

\preprint{APS/123-QED}

\title{Quantum curvature fluctuations and the cosmological constant in a single plaquette quantum gravity model}

\author{Samuel Blitz}%
 \email{blitz@math.muni.cz}
\affiliation{
 Department of Mathematics and Statistics, Masaryk University\\ Building 08, Kotlářská 2,
Brno, Czech Republic 61137
}%

\author{Shahn Majid}
\email{s.majid@qmul.ac.uk}
\affiliation{School of Mathematical Sciences,\\
Queen Mary University of London, London E1 4NS
}%

\date{\today}

\begin{abstract}

Understanding the microscopic behavior of spacetime, in particular quantum uncertainty in the Ricci scalar,  is critical for developing a theory of quantum gravity and perhaps solving the cosmological constant problem. To test this, we compute this quantity for a simple but exact discrete quantum gravity model based on a single plaquette of spacetime. Our results confirm initial speculations of Wheeler from 1955 in finding a UV divergence in the quantum uncertainty. We further show that this behavior is stable under renormalization, but potentially unstable with the introduction of a cosmological constant, suggesting that a bare cosmological constant is ruled out.


\end{abstract}

\maketitle


\section{Introduction}
Perhaps the most longstanding problem in fundamental physics is the quantization of gravity. It is hoped that whatever the ``correct" theory of quantum gravity might be, such a theory would enable us to solve a variety of open problems in physics. Of interest in this letter is the \textit{cosmological constant problem}: that the naively-calculated vacuum energy density is 50-122 orders of magnitude larger than the observed dark energy density. Among other things, this disagreement implies either an incredibly fine-tuned cancellation occuring or a subtlty in the theory that has not been taken into account~\cite{Martin2012}.

In recent works by Carlip~\cite{Carlip2019}, Wang and Unruh~\cite{wang2019}, and Wang~\cite{wang2024}, it is suggested that the problem disappears when one considers the gravitationally attractive nature of microscopic curvature fluctuations. These were  heuristically discussed in \cite{geons,Dewitt,MTW} using  Wheeler's ``spacetime foam" with magnitude proposed  in these works as being of order $\Delta R \sim \ell_p / L^3$, where $\ell_p$ is the Planck length and $L$ is the size of the region under consideration. With this estimate, ultra-local (i.e. Planckian)  quantum gravity fluctuations contribute to an effective attractive energy density which overwhelm the (by comparison) paltry repulsive vacuum energy density associated with particle fields.

The idea in ~\cite{Carlip2019,wang2019,wang2024} is that, as the fluctuations are in the spacetime geometry itself, spacetime should be treated as highly inhomogeneous. So, at an intermediate length scale well-above the Planck scale but well-below other relevant scales, we may locally model the spacetime fluctuations as an effective stress-energy source within a \textit{local} Friedmann equation given by
\begin{align}
\frac{\ddot{a}}{a} = -\frac{4 \pi G}{3} (-2 \rho_{SM} + (1+3w) \rho_{QG})\,,
\end{align}
where $\rho_{SM}$ is the average vacuum energy density of the Standard Model calculated in the typical way and diverging as the UV cutoff goes to infinity (with the usual effective equation of state), $\rho_{QG}$ is the local average energy density associated with curvature fluctuations, and $w$ characterizes the effective equation of state of the curvature fluctuations. Note that the local Friedmann equation does not require that the scale factor $a$ is spatially constant as the spatial derivatives cancel out. Moreover,  $w > -1/3$ because the gravitational fluctuations are always attractive.  Hence, when $\rho_{QG} \gg \rho_{SM}$, solutions to the local Friedmann equation become oscillatory rather than exponential. The idea is that with each small region of spacetime rapidly oscillating independently according to the local Friedmann equation, the scale factor vanishes when averaged over a region many orders of magnitude larger than a single domain.  Rigorously justifying this otherwise-intuitive behavior is quite challenging, see~\cite{Buchert,Wiltshire,CELU,GreenWald} for more details on this so-called ``averaging problem". This gives an argument for, absent any additional higher order parametric resonance effects,  a vanishing Hubble constant and thus a vanishing macroscopic cosmological constant. 

Such a proposal turns around the cosmological constant problem from trying to construct a model that reduces a large ``bare'' cosmological constant to the small observed value, to constructing models that deviate slightly from a value that would otherwise be zero. This also dovetails with recent results from the DESI collaboration~\cite{desi} which suggest that the cosmological constant is better fit by a time-dependent field rather than a constant intrinsic to spacetime itself. If this is the case, then the observed expansion cannot be driven by a cosmological constant term in Einstein's field equations, and thus we would expect $\Lambda = 0$ on the macroscopic scale.  

As the implication of such arguments are significant, the primary purpose of this letter is to revisit the previous heuristics about quantum uncertainty in the curvature by means of computations within an exact toy quantum gravity model. The model we use is the quantum Lorentzian geometry of a two-dimensional ``plaquette" of spacetime $(\mathbb{Z}_2)^2$ in \cite{Ma:sq}. In principle, we are most interested in the behavior of a small spacetime volume, modelled as a single $3+1$ dimensional hypercube of spacetime $(\Z_2)^4$.  Both dimensional reduction~\cite{ambjorn2001,freidel2006} and the use of  lattice models~\cite{Hamber2009} are established methods to examine quantum gravity. Moreover, the particular approach we use is noncommutative (or quantum)  Riemannian geometry as in \cite{MaTao,Ma:sq,ArgMa,LirMa} applied in the case of a bidirected graph, and is  distinct from other approaches such as spin networks~\cite{RovelliSmolin1995} or dynamical triangulations~\cite{Loll2020}. The Gauss-Bonnet theorem making 2D gravity topological does not necessarily apply in noncommutative geometry and indeed the single plaquette model has dynamical modes. Hence, we believe that we may still obtain qualitative information about the behavior of this system by instead working with the more computable $(\Z_2)^2$ case. As a graph geometry, it has no external legs and hence can be considered as without boundary.

This approach replaces the functions on a manifold by the algebra $A$ of functions on the set of vertices, the `1-forms' by a vector space $\Omega^1$ with basis $\{\omega_{x\to y}\}$ labelled by the arrows, with exterior derivative $\extd: A\to \Omega^1$ the finite differences $\extd f=\sum_{x\to y} (f(y)-f(x))\omega_{x\to y}$. The main difference from regular lattice theory is that we multiply 1-forms and functions by $f\omega_{x\to y}=f(x)\omega_{x\to y}$ and $\omega_{x\to y}f= f(y) \omega_{x\to y}$  to have an exact Leibniz rule,  but at the price that functions and 1-forms do not generally commute. A metric is a nondegenerate element $g\in \Omega^1\tens_A\Omega^1$ and in the graph case amounts to assigning a non-zero real `square length' to every arrow. There is no quantisation or deformation parameter in the model as this is simply an exact finite noncommutative geometry. Instead, the metric values and cutoffs that we will impose on them will provide physical scales. We further require  a notion of symmetry on the metric by requiring that the square-lengths are the same for arrows in opposite directions. Thus, for the Lorentzian square\cite{Ma:sq}, the metric data is
{\ } \vskip-15pt
\begin{figure}[H]
\begin{center}
\includegraphics[scale=0.2]{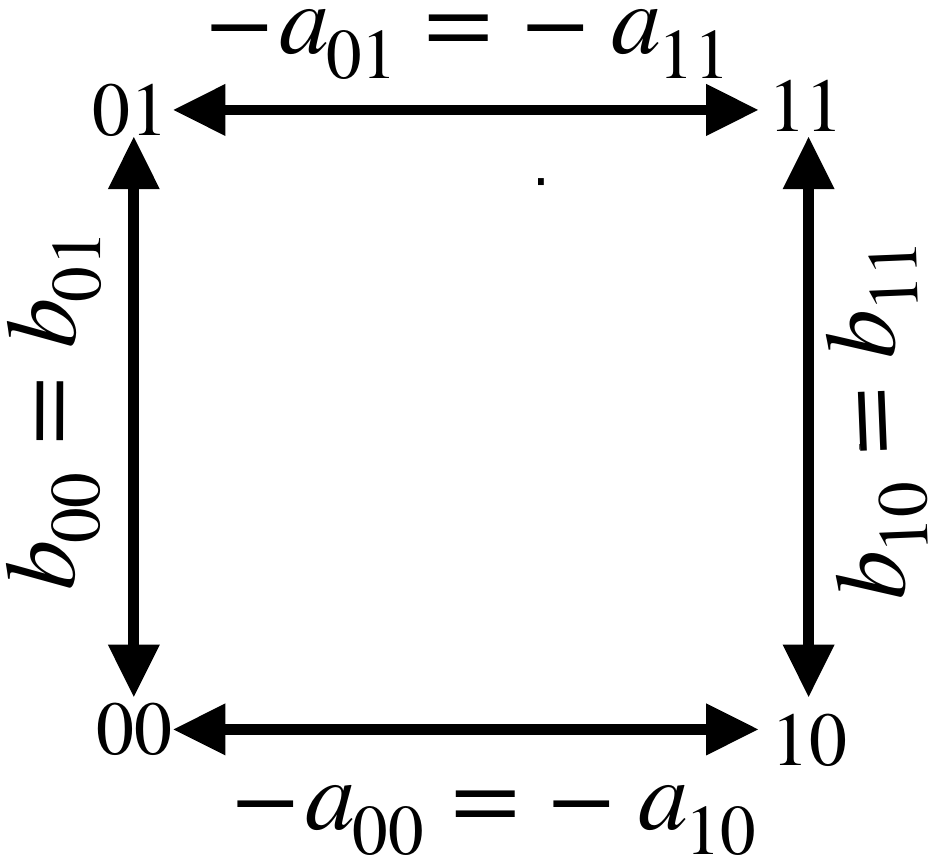}
\end{center}
\end{figure}
 \vskip-7pt
\noindent with horizontal edges assigned a spacelike or negative value and vertical edges a timelike or positive one. We labelled the vertices as elements of the group $\Z_2\times\Z_2$. With respect to the latter,  $\Omega^1$ has a 2-dimensional basis over $A$ of left-invariant 1-forms $e_1, e_2$ and  $g=-a e_1\tens e_1+ be_2\tens e_2$, with  the metric data now appearing as certain positive metric coefficients  $a,b\in A$ valued as shown.  There are also higher differential forms with $e_1,e_2$  forming a Grassmann algebra under the wedge product $\wedge$ and killed by the exterior derivative $\extd$.
 
Given a metric, one then solves for a  torsion-free metric-compatible ``quantum Levi-Civita'' connection $\nabla:\Omega^1\to \Omega^1\tens_A\Omega^1$, finding in the present case\cite{Ma:sq}  a 1-parameter family labelled by a phase $q=e^{\imath\theta}$. (This phase is a `purely quantum' feature of the geometry since classically the Levi-Civita connection is unique.) Here the left most factor of $\nabla$ can be evaluated against a vector field (in the noncommutative geometry sense) to get a covariant derivative along it). From $\nabla$, there is a canonical Riemann curvature $R_\nabla=(\extd\tens\id-\id\wedge\nabla)\nabla$ 
as a 2-form-valued operator on one forms. Viewing the 2-form-values as  antisymmetric tensors, one can then trace to get the Ricci scalar curvature $R$.  Evaluating the expression for this found in \cite{Ma:sq} at the four points  gives
\begin{align}\label{Rab} R(00)&={1\over a_{00}b_{00}}\left( \frac{a_{00}^2}{a_{01}}-a_{00}+({1+q^{-1}\over 2}) (b_{00}-b_{10})\right) ,\nonumber\\ 
R(01)&=-{1\over a_{01}b_{00}}\left( \frac{ b_{00}^2}{b_{10}}-b_{00} +({1+q\over 2}) (a_{00}-a_{01})\right) ,\nonumber\\ 
R(10)&=-{1\over a_{00}b_{10}}\left( {b_{10}^2\over b_{00}}- b_{10} -({1+q\over 2})(a_{00}-a_{01})\right),\nonumber\\ 
R(11)&={1\over a_{01}b_{10}}\left(\frac{ a_{01}^2}{a_{00}}-a_{01}-({1+q^{-1}\over 2}) (b_{00}-b_{10}) \right).\end{align}
For the Einstein-Hilbert action we sum $R$ over the four points but with weight function $ab$ (this is the negative determinant of the matrix of $g$ in our basis). This is chosen as in \cite{Ma:sq} to eliminate the dependence on $\theta$ and results, as found there, in
\begin{equation} S_g= (a_{00}-a_{01})^2({1\over a_{00}}+{1\over a_{01}})-(b_{00}-b_{10})^2({1\over b_{00}}+{1\over b_{10}}).  \end{equation}
Next, for calculations, it is convenient to   $\Z_2\times\Z_2$-Fourier transform the metric coefficients, which amounts to
\begin{align}a_{00}=a_{10}=k_0(1+k),\quad a_{01}=a_{11}=k_0(1-k),\nonumber\\
b_{00}=b_{01}=\ell_0(1+\ell),\quad b_{10}=b_{11}=\ell_0(1-\ell),\end{align}
in terms of  $k_0,\ell_0> 0$ the average horizontal, resp. vertical, edge values and $k,\ell\in(-1,1)$ the horizontal resp. vertical relative fluctuations. In these field momentum variables, the action becomes
\begin{equation} S_g=k_0\alpha(k)-\ell_0\alpha(\ell),\quad \alpha(k)={8 k^2\over 1-k^2}.\end{equation}
If we control UV and IR divergences by limits $\varepsilon < k_0, \ell_0 < \mathcal{L}$ then the regulated quantum gravity partition function is given by
\begin{equation} Z = 4 \int_{-1}^{1} \int_{-1}^{1}  \int_{\varepsilon}^{\mathcal{L}} \int_{\varepsilon}^{\mathcal{L}}  \extd k\, \extd \ell \, \extd k_0 \, \extd \ell_0 \, k_0 \ell_0 \,e^{iS_g / G}\,.\end{equation}
For any quantum operator $\langle \mathcal{O} \rangle$, we can compute its corresponding regulated correlation function according to
\begin{equation}\label{vevO} \langle \mathcal{O} \rangle = \frac{4 \int_{-1}^{1} \int_{-1}^{1}  \int_{\varepsilon}^{\mathcal{L}} \int_{\varepsilon}^{\mathcal{L}}  \extd k\, \extd \ell \, \extd k_0 \, \extd \ell_0 \, k_0 \ell_0 \mathcal{O} \,e^{iS_g / G}}{Z}\,.\end{equation}
Notice that the $\ell,\ell_0$ theory is the complex conjugate of the $k,k_0$ theory. Hence it is enough to work just with the latter and obtain the other side by complex conjugation. Also note that $Z$ is real and differentiating with respect to the coupling constant brings down $\imath S_g$, hence the expectation value of $S_g$ is necessarily imaginary. So we expect and will find that the expectation of the averaged Ricci scalar is similarly imaginary.

Next, we note that observables which are odd in $k$ have vanishing correlation functions and if we work with even functions of $k$, we can restrict to $k\in [0,1)$. We can then change variables to $\alpha(k)\in [0,\infty)$ in place of $k=\sqrt{\alpha/(\alpha+8)}$. The $k,k_0$ factor of $Z$ can now be computed up to constant factors as \cite{Ma:sq}
\begin{align}\label{Z1} Z_1&=\int_0^\infty {\extd \alpha \over \alpha^{1\over 2}(\alpha+8 )^{3\over 2}} \int_{\eps}^{\L} \extd k_0\,  k_0 e^{\imath k_0 \alpha/G}\nonumber\\
&= \int_0^\infty {\extd \alpha \over \alpha^{1\over 2}(\alpha+8 )^{3\over 2}} {\extd\over\extd \alpha}\left({e^{\imath  \L \alpha /G}-e^{\imath\eps \alpha/G}\over \alpha}\right)\end{align}
and similar expressions with $\mathcal O$ inserted when computing expectation values of functions of $k_0,k$.

\section{The Averaged Ricci Scalar}\label{average-section}
For the purposes of this article, we are interested in the behavior of the vertex-averaged, $\theta$-averaged Ricci scalar  to reflect what is seen at much larger scales.  Averaging (\ref{Rab}) over $\theta$ with uniform measure and  converting to field momentum variables, we obtain
\begin{align}
R(00) &= \frac{2k}{1-k} \frac{1}{\ell_0(1+\ell)}, \nonumber\\
R(01) &= -\frac{2}{k_0 (1-k)} \frac{\ell}{1-\ell},\nonumber \\
R(10) &=  \frac{2}{k_0 (1+k)} \frac{\ell}{1+\ell},\nonumber\\
R(11) &= - \frac{2k}{1+k} \frac{1}{\ell_0 (1- \ell)}\,.
\end{align}
Now note that because expectation values (\ref{vevO}) of odd functions of $k$ or $\ell$ vanish, we may write that
\begin{align}\label{vevk}
\left \langle \frac{k}{1-k} \right \rangle &= \left \langle \frac{k^2}{1-k^2} \right \rangle = \frac{1}{8} \langle \alpha \rangle, \nonumber\\
\left \langle \frac{k}{1+k} \right \rangle &=  -\left \langle \frac{k^2}{1-k^2} \right \rangle = -\frac{1}{8} \langle \alpha \rangle, \nonumber\\
\left \langle \frac{1}{k_0(1-k)} \right \rangle &=  \left \langle \frac{1}{k_0 (1-k^2)} \right \rangle = \left \langle \frac{8+\alpha}{8 k_0} \right \rangle,\nonumber \\
\left \langle \frac{1}{k_0(1+k)} \right \rangle &=  \left \langle \frac{1}{k_0 (1-k^2)} \right \rangle = \left \langle \frac{8+\alpha}{8 k_0} \right \rangle \,.
\end{align}
Defining $R_{av} = \tfrac{1}{4}(R(00) + R(01) + R(10)+ R(11))$, we find that
\begin{align}\label{Rav}
\langle R_{av} \rangle 
&= \tfrac{1}{64} \langle \alpha \rangle \overline{\left \langle \frac{8+\alpha}{k_0} \right \rangle} - \tfrac{1}{64} \overline{\langle \alpha \rangle} \left \langle \frac{8+\alpha}{k_0} \right \rangle \nonumber \\
&= \tfrac{i}{32} \operatorname{Im}\left( \langle \alpha \rangle \overline{\left \langle \frac{8+\alpha}{k_0} \right \rangle} \right)\,.
\end{align}
We use parallel formulae to (\ref{vevk}) for $l_0,l$ but recall that these enter with a complex conjugation. These bring their own $1/8$ factor resulting in the $1/64$. 

A similar calculation yields the variance of the averaged Ricci scalar:
\begin{equation} \langle R_{av}^2 \rangle = { \operatorname{Re}\over 1024} \left(\langle \alpha^2 \rangle \overline{\langle \tfrac{(\alpha+4)(\alpha+8)}{k_0^2} \rangle} + 4 \langle \alpha \rangle \overline{\langle \tfrac{\alpha (8+\alpha)}{k_0^2} \rangle} \right)\,. \end{equation}
An interesting observation is that while $\langle R_{av} \rangle$ is always imaginary as expected, $\langle R_{av}^2 \rangle$ is always real. This suggests that it is the variance of the Ricci scalar, rather than the Ricci scalar itself, which is a physical observable.

\medskip

These correlators still depend on the limits $\mathcal{L}$ and $\mathcal{\eps}$ in (\ref{Z1}) and we look at them in a large $\mathcal{L}$ and small $\varepsilon$ limit. However, even then, the necessary integrals are non-trivial.  In some of these cases, named integrals are obtainable (in the form of Meijer-G functions), but series expansions for these special functions are non-trivial. As such, we used the inverse equation solver RIES~\cite{ries} to determine the exact form of leading and subleading terms in powers of $\mathcal{L}$ from numerics. When this method was unusable, we instead approximated the highly-oscillatory integrands via Fourier methods described in~\cite[Section 1]{milovanovic2013}. Finally, when even these methods did not produce convergent integrals, we approximated the integrands using piecewise-stitched Taylor series approximations.

Implementing these approximations, we may compute the regulated partition function $Z := |Z_1|^2$, finding 
\begin{equation}Z_1 \approx \tfrac{1}{3} (1+i) \sqrt{\pi G \mathcal{L}^3} + \tfrac{1}{16}(1-i) \sqrt{\pi G^3 \mathcal{L}} \,,\end{equation}
to subleading order in $\mathcal{L}$. Here and going forward, we expand first about $\varepsilon \rightarrow 0$, keeping only the lowest order term, and then expand about $\mathcal{L} \rightarrow \infty$.

We can also approximate several correlators:
\begin{align}
    \langle k_0 \rangle &\approx \frac{3 \mathcal{L}}{5} + \frac{3iG}{160}, \nonumber\\
    \langle \alpha \rangle &\approx \frac{3i G}{2 \mathcal{L}} + \frac{27(1-i)}{35} \sqrt{\frac{G^3}{\mathcal{L}^3}} ,\nonumber\\
    \langle \alpha^2 \rangle &\approx \frac{6(i-1)}{\sqrt{\pi}} \sqrt{\frac{G^3}{\mathcal{L}^3}} + \frac{9G^2}{4 \mathcal{L}^2},\nonumber \\
    \left \langle \frac{8+\alpha}{k_0} \right \rangle &\approx \frac{24}{\mathcal{L}} - \frac{3(1+i)}{\sqrt{\pi}} \sqrt{\frac{G}{\mathcal{L}^3}} ,\nonumber\\
    \langle \tfrac{(\alpha + 4)(\alpha+8)}{k_0^2} \rangle &\approx \frac{24(1+i)}{\sqrt{\pi}} \sqrt{\frac{G}{\varepsilon^2 \mathcal{L}^3}} + \frac{9(i-1)}{2\sqrt{\pi}} \sqrt{\frac{G^3}{\varepsilon^2 \mathcal{L}^5}},\nonumber \\
\langle \tfrac{\alpha(\alpha+8)}{k_0^2} \rangle &\approx \frac{24(1+i)}{\sqrt{\pi}} \sqrt{\frac{G}{\varepsilon^2 \mathcal{L}^3}} + \frac{9(i-1)}{2\sqrt{\pi}} \sqrt{\frac{G^3}{\varepsilon^2 \mathcal{L}^5}}\,.
\end{align}
Combining these computations, we have that
\begin{equation}\langle R_{av} \rangle \approx \frac{9iG}{8\mathcal{L}^2} - \frac{9i(144 \sqrt{\pi} + 35)}{2240 \sqrt{\pi}} \sqrt{\frac{G^3}{\mathcal{L}^5}}\end{equation}
and
\begin{equation}\langle R_{av}^2 \rangle  \approx \frac{9}{64\sqrt{\pi}} \sqrt{\frac{G^3}{\varepsilon^2 \mathcal{L}^5}} + \frac{81}{1024 \sqrt{\pi}}\sqrt{\frac{G^5}{\varepsilon^2 \mathcal{L}^7}} \,. \end{equation}
Indeed, keeping only the leading term in $\varepsilon$, we may write the uncertainty of the averaged Ricci scalar:
\begin{equation}\Delta R_{av} \approx \frac{3 }{8 \pi^{1/4}} \frac{G^{3/4}}{ \varepsilon^{1/2} \mathcal{L}^{5/4}}\,.\end{equation}
The approximations used here are robust under variation of the interpolations and approximations. We thus have an interesting result: there is a UV divergence in the gravitational fluctuations  for a fixed IR cutoff $\mathcal L$ and a vanishing averaged Ricci scalar as $\mathcal L\to\infty$.

There is risk, however, that this behavior is qualitatively changed under renormalization. We resolve this question in the next section.

\section{Renormalization}

The quantities in previous sections were in physical units and for clarity we now label them explicitly as such with the superfix {\em phys}. We then work in  Planckian units in which we redefine 
\begin{align} \mathcal{L} &= {\mathcal{L}^{phys}\over G},\quad \varepsilon = {\varepsilon^{phys}\over G},\quad  \langle k_0 \rangle_\mathcal{L} ={ \langle k_0^{phys} \rangle\over G},\nonumber\\
 \langle R_{av} \rangle_\mathcal{L} &= G \langle R_{av}^{phys} \rangle,\quad  \langle R_{av}^2 \rangle_\mathcal{L} = G^2 \langle (R_{av}^2)^{phys} \rangle,\end{align}
 where we also write the regulator explicitly in the expectation values. In terms of the rescaled $\mathcal L,\eps$, we now have
\begin{align}
\langle k_0 \rangle_{\mathcal{L}} \approx& \frac{3 \mathcal{L}}{5} + \frac{3i}{160},\nonumber \\
\langle R_{av} \rangle_\mathcal{L} \approx& \frac{9i}{8 \mathcal{L}^2}  - \frac{9i(144 \sqrt{\pi} + 35)}{2240 \sqrt{\pi} \mathcal{L}^{5/2}} ,\nonumber \\
\langle R_{av}^2 \rangle_{\mathcal{L}} \approx& \frac{9}{64\sqrt{\pi}} \sqrt{\frac{1}{\varepsilon^2 \mathcal{L}^5}} + \frac{81}{1024 \sqrt{\pi}}\sqrt{\frac{1}{\varepsilon^2 \mathcal{L}^7}}\,.
\end{align}

Our approach to the renormalization is to solve for the IR regulator $\mathcal L$ in terms of a fixed $\langle k_0^{phys} \rangle$. Then 
\begin{equation}\langle R^{phys}_{av} \rangle = \langle R_{av} \rangle_{\mathcal{L}} \frac{\langle k_0 \rangle_{\mathcal{L}}}{\langle k_0^{phys} \rangle}= \frac{81iG}{200\langle k_0^{phys} \rangle^2} + \mathcal{O}(G^{3/2})\, , \end{equation}
where we replaced $\mathcal L$ in terms of $\langle k_0^{phys} \rangle$. As expected, any non-vanishing behavior in the averaged Ricci scalar is purely quantum in nature. Similarly replacing $\mathcal L$ in $\langle{R_{av}^2}\rangle_\mathcal{L}$ and converting back to physical quantitites, we find that the renormalized variance of the Ricci scalar takes the form 
\begin{equation}\langle (R^2_{av})^{phys} \rangle = \frac{81 \sqrt{15}}{8000 \sqrt{\pi}} \sqrt{\frac{G^3}{(\varepsilon^{phys})^2 \langle k_0^{phys} \rangle^5}} + \mathcal{O}(G^{5/2})\, .\end{equation}
 We see that this is a genuine quantum effect and that the behavior (with an adjustment to the coefficient) remains stable under renormalization, namely, a UV divergence still arises as in Section~\ref{average-section} in terms of renormalised quantities.

\section{Physical Interpretation} \label{sec:physics}
The construction in Section~\ref{average-section} provides a more detailed account of the behavior leading to large fluctuations than in~\cite{geons,Dewitt,MTW}. There, a fixed-size region of spacetime is specified and fluctuations are viewed as resonances in that volume, with the magnitude of the fluctuations depending on the scale of the spacetime volume. However, while this heuristic description is a useful starting point, the notion that one begins by considering a fixed volume spacetime is contrary to the spirit~\cite{Smolin2006} of a background-independent theory of quantum gravity. The model provided here using quantum geometry gives a properly background-independent approach that leads to curvature fluctuations, as we have not specified any fixed metric. Instead, we have merely provided a graph structure, which is akin to providing a manifold or differential structure rather than a metric.

Physically, we view this model as approximating the behavior of a ``small'' subset of the spacetime manifold prior to specification of the metric. By then imposing a quantum geometry on the manifold and minimum and maximum cutoffs on the sizes of the dimensions of that subset, we find that the approximate size of the subset is determined by the expectation value $\langle k_0 \rangle$, inducing a metric. Unlike the heuristic in~\cite{geons,Dewitt,MTW}, the magnitude of curvature fluctuations depends also on the \textit{minimum} size of the region of spacetime. One way to see this is that derivatives of metric functions depend both on the discrete difference in the metric functions and inversely on the metric functions themselves. If the metric functions are allowed to approach zero, fluctuations can diverge.

Nonetheless, the behavior of this model still reproduces the hierarchy of energy densities required by~\cite{wang2024} so long as appropriately physical regulators are introduced.  As we intend to model only a ``unit'' of quantum spacetime in a way that is insensitive to matter field fluctuations, the IR cutoff should be \textit{shorter} than the scale where Standard Model fields can see the fluctuations. Otherwise, we would already have seen the effects of the discretisation in the Standard Model.

Consider, for example, the conservative case where we choose a UV cutoff at the Planck length and an IR cutoff of $\mathcal{L} = (2 \times 10^{-21} \text{ m})^2$, corresponding to $100 \text{ TeV}$, as suggested by~\cite{Carmona_2001}. In that case, using the Einstein field equations we may approximate the equivalent energy density of the fluctuations according to
\begin{equation}\rho_{QG} = \frac{c^4 \Delta R_{av}}{8 \pi G}\,.\end{equation}
So, we find that $\rho_{QG} \sim 10^{39} \text{ GeV}^4$, compared with the estimated~\cite{Martin_2012} QFT vacuum energy of $\rho_{SM} \sim 10^8 \text{ GeV}^4$. In fact, this hierarchy is preserved for any physically reasonable choices of cutoffs. Certainly, a reasonable UV cutoff should be at or below the Planck length, and the IR cutoff should correspond to a length scale above which we measure no quantum gravitational effects, i.e. perhaps at the scale of the Standard Model or shorter. Had we instead chosen the GUT scale $ L_{\text{GUT}} \approx 10^{-32} \text{ m}$, the energy gap would have been even more extreme.

These estimations support the critical assertion in~\cite{wang2024} that $\rho_{QG} \gg \rho_{SM}$, implying that for a given region of spacetime the size of the Lorentzian square examined here, the scale factor is oscillatory rather than exponential. Furthermore, taking the IR length scale at $2 \times 10^{-21} \text{ m}$, we find from $\langle k_0 \rangle$ that the average size of the spacetime domain is on the order of $10^{-21} \text{ m}$, meaning that the wavelength of the fluctuations are of the same order. Thus, averaging over even a cubic centimeter of space would mean summing the fluctuations of $10^{56}$ microscopic domains of spacetime, clearly sufficiently many to average out fluctuations~\cite{Carlip2019}. Similarly, the time scale of these fluctuations is on the order of $10^{-30} \text{ s}$, whereas the smallest time yet measured~\cite{zepto} is on the order of $10^{-19} \text{ s}$, meaning that we may average over $10^{10}$ spacetime domains temporally---again, clearly sufficient to average out the fluctuations.

To see an explicit example of the kind of averaging considered here, see~\cite[Section 5e]{wang2024}. Essentially, one may consider the volume of a given macroscopic region of spacetime by considering the volume of each Planckian subregion (whose geometry is rapidly fluctuating); by adding up the contributions of each subregion and noting that amplitudes of each oscillation do not depend on time, each spatial slice of the macroscopic spacetime volume is shown to be constant.  The implication is that, absent any other effects aside from the quantum geometric effects and the Standard Model vacuum fluctuations, the scale factor would be measured to be constant over any macroscopic scale.

While the above computation should be treated as qualitative, it does show that the argument used in~\cite{Carlip2019,wang2019,wang2024}---that the gravitational fluctuations dominate the Standard Model energy density when we consider the quantum gravity regime---is robust across many quantum gravitational models, even those as simple as a $1+1$d discrete quantum system as discussed above.

\section{A Perturbative Cosmological Constant}
In~\cite{wang2024}, it was also assumed that the intrinsic macroscopic cosmological constant of the spacetime vanishes, and that any cosmological constant must arise from quantum behavior. Interestingly, the model constructed in Section~\ref{average-section} also provides a hint that this constraint is reasonable. To see this, we consider the behavior of curvature correlators in the presence of a small cosmological constant (such that we may keep only terms linear in $\Lambda$). To that end, we perturb the action and the corresponding correlators merely by writing 
\begin{equation}\scalebox{0.9}{$Z_\Lambda = 4 \int_{-1}^{1} \int_{-1}^{1}  \int_{\varepsilon}^{\mathcal{L}} \int_{\varepsilon}^{\mathcal{L}}  \extd k\, \extd \ell \, \extd k_0 \, \extd \ell_0 \, k_0 \ell_0 \operatorname{exp}\left[\frac{i}{G} (S_g - 4 \Lambda k_0 \ell_0) \right]$}\end{equation}
and expanding in powers of $\Lambda$, keeping only the lowest order term:
\begin{equation}Z_{\Lambda} \approx \left(1 - \tfrac{4i \Lambda}{G}| \langle k_0 \rangle_0|^2 \right) Z_0\,. \end{equation}

Similarly, to leading order in $\Lambda$ correlators can be computed as
\begin{align}
\langle \mathcal{O} \rangle_{\Lambda} &\approx \langle \mathcal{O} \rangle_0  + \frac{\partial \langle \mathcal{O} \rangle_{\Lambda}}{\partial \Lambda}\Big|_{\Lambda = 0} \Lambda \nonumber\\
&= \langle \mathcal{O} {\rangle}_{0} -\tfrac{4i \Lambda}{G} \big( \langle k_0 \ell_0 \mathcal{O} \rangle_{0} - \langle k_0 \ell_0 \rangle_{0} \langle \mathcal{O} \rangle_{0} \big) \nonumber\\
&= \left(1 + \tfrac{4i \Lambda}{G} |\langle k_0 \rangle_{0}|^2 \right) \langle \mathcal{O} {\rangle}_{0} -\tfrac{4i \Lambda}{G} \langle k_0 \ell_0 \mathcal{O} \rangle_{0}\,,
\end{align}
where $\langle \mathcal{O} \rangle_0$ is the correlator in the theory with vanishing cosmological constant.

As in Section~\ref{average-section}, we may compute the necessary correlation functions:
\begin{align}
    \langle k_0 \ell_0 R_{av} \rangle &\approx \frac{iG}{8} + \frac{27i}{2240} \sqrt{\frac{G^5}{\mathcal{L}^3}},\nonumber \\
    \langle k_0 \ell_0 R_{av}^2 \rangle &\approx (0.0079 + 0.0264 \ln G/\epsilon) (G/\mathcal{L})^{3/2}
\end{align}

Using these, we find that
\begin{equation}\langle R_{av} \rangle_{\Lambda} \approx - \frac{28}{25} \Lambda + 0.95 \Lambda \sqrt{\frac{G}{\mathcal{L}}}\,, \end{equation}
and
\begin{equation}
\langle R_{av}^2 \rangle_{\Lambda} \approx \frac{9}{64\sqrt{\pi}} \sqrt{\frac{G^3}{\varepsilon^2 \mathcal{L}^5}} + \frac{81i \Lambda}{400 \sqrt{\pi}} \sqrt{\frac{G}{\varepsilon^2 \mathcal{L}}}.
\end{equation}
Note that, as expected, the introduction of a cosmological constant induces a non-vanishing Ricci scalar in the large $\mathcal L$ limit.  (Caution is advised when comparing this expectation value to the Ricci scalar in a classical vacuum universe: the dynamics here are qualitatively different, despite arising from ``quantizing" a classical action.) However, $\langle R_{av}^2 \rangle_{\Lambda}$ is modified at \textit{lower} order in $1/\mathcal{L}$, which suggests instability in the fluctuations that depend on the cosmological constant. We speculate that this instability can provide a potential reason for enforcing a vanishing cosmological constant: non-smooth behavior in the magnitude of curvature fluctuations seems unphysical.

We caution, however, that $\langle R_{av}^2 \rangle_{\Lambda}$ cannot be directly interpreted as the correlator is no longer real, nor do  we have a fully non-perturbative functional description of $\langle R^2_{av} \rangle_{\Lambda}$.

\section{Conclusion}
In this letter, we have used an explicit quantum gravity model based on quantum geometry to compute spacetime curvature fluctations at the ultra-microscopic level. Our results are generally consistent with heuristic arguments regarding the magnitude of spacetime fluctuations and, as such, suggest that a recently proposed resolution~\cite{Carlip2019,wang2019,wang2024} of the cosmological constant problem has at least some theoretical merit. Furthermore, the implication of a novel instability in the vacuum in the presence of a perturbatively small cosmological constant leads to additional speculation that, perhaps, the intrinsic cosmological constant must vanish, consistent with recent observations from the DESI collaboration~\cite{desi}.  We also found that $\langle R_{av}\rangle$ is imaginary, as expected for a Lorentzian model, but disappears for large $\mathcal L$.

This work is merely a first step in examining the kinds of physics that can be extracted from baby quantum gravity models based on  quantum geometry.  The case of Lorentzian $(\Z_2)^4$ should be looked at but is currently too hard to compute, while other currently computable quantum gravity models using quantum geometry~\cite{MaTao,ArgMa,LirMa} are Euclidean rather than Lorentzian. While our application has been explicitly tied to small scales, one could also use graphs with increasingly many points to model spacetime at large, while scaling all metric `square-lengths' so that the vertices become in some sense `closer' together. Depending on how the scaling does, the noncommutativity of the differential forms and functions can still be present in the limit, with some of the freedom appearing as a deformation parameter (this parameter would not, however, be expected to be $\hbar$ but the Planck scale if the original discreteness was a Planck scale effect). The fact that quantum Riemannian geometry can be specialised to both the graph case and the classical (as well as noncommutative) continuum case means that these limits in principle make sense, although some quantities could still diverge in the limit. How to formulate a Lorentzian graph geometry in general, however, remains unclear,  but it may be possible to borrow some ideas here from causal set theory. Doing quantum gravity on such large graphs, although currently too hard to compute, could then be attempted in lieu of continuum quantum gravity. In principle, one could also sum over all bidirected graphs with a fixed number of vertices, which would be akin to considering all possible differential structures on a given topological space, which is a step beyond quantum gravity as normally conceived. It would also be important to introduce particles, and in our specific application above to quantify the relative difference in energies associated with gravitational fluctuations and particle fluctuations.

{\ }

\bigskip

\section*{Acknowledgements}
SB was supported by the Operational Programme Research Development and Education Project No. CZ.02.01.01/00/22-010/0007541.

\bibliography{QRSbib}

\end{document}